\RequirePackage{fix-cm}
\documentclass[smallextended]{svjour3}       
\smartqed  

\usepackage{graphicx,mathptmx,color,subfigure}      
\usepackage{amsmath,amssymb,bm}
\newcommand{\openone}{\leavevmode\hbox{\small1\normalsize\kern-.33em1}} 
\newcommand{\Gal}[1]{\mathbb{F}_{#1}}
\newcommand{\Tr}{\mathop{\mathrm{Tr}}\nolimits} 
\newcommand{\tr}{\mathop{\mathrm{tr}}\nolimits} 
\newcommand{\sugg}[1]{{#1}}
\newcommand{\fact}{\mathop{\mathrm{fact}}\nolimits} 

\journalname{Quantum Information Processing}

\begin{document}

\title{Discrete phase-space structures and Wigner functions for $N$~qubits}


\author{C. Mu\~{n}oz$^{1}$   \and A. B. Klimov$^{1}$ \and
L. S\'anchez-Soto$^{2,3}$}


\institute{$^{1}$  Departamento de F\'{\i}sica, 
Universidad de Guadalajara, 44420~Guadalajara, 
Jalisco, Mexico \\
$^{2}$ Departamento de \'Optica, Facultad de F\'{\i}sica, 
Universidad Complutense, 28040~Madrid,  Spain\\ 
$^{3}$ Max-Planck-Institut f\"ur die Physik des Lichts,
Staudtstra\ss e 2, 91058 Erlangen, Germany} 

\date{Received: \today / Accepted: date}

\maketitle

\begin{abstract}
\sugg{We further elaborate on}  a phase-space picture for a
system of $N$ qubits and explore the structures compatible with the notion
of unbiasedness. These consist of bundles of discrete curves satisfying
certain additional properties and different entanglement properties. We
discuss the construction of discrete covariant Wigner functions for these
bundles and provide several illuminating examples.
\end{abstract}




\institute{$^{1}$  Departamento de F\'{\i}sica, 
Universidad de Guadalajara, 44420~Guadalajara, 
Jalisco, Mexico \\
$^{2}$ Departamento de \'Optica, Facultad de F\'{\i}sica, 
Universidad Complutense, 28040~Madrid,  Spain\\ 
$^{3}$ Max-Planck-Institut f\"ur die Physik des Lichts,
Staudtstra\ss e 2, 91058 Erlangen, Germany}

\section{Introduction}

Phase-space methods offer the remarkable advantage that quantum
mechanics appear as similar as possible as a classical statistical
theory, by avoiding the operator
formalism~\cite{Schroek:1996aa,Schleich:2001aa,QMPS:2005aa}.

\sugg{The relevant role of  discrete quantum systems, which live
  in a $d$-dimensional Hilbert space, was early anticipated by
  Weyl~\cite{Weyl:1928aa}. The  related problem of generalizing
  the Wigner function to these finite systems has a long history. 
A plausible approach was taken by Hannay
and Berry~\cite{Hannay:1980aa}, considering a phase space
constrained to admit only periodic probability distributions,
which implies that the corresponding manifold is effectively a
$2d \times 2d$-dimensional torus. Other surrogates using a $2d \times
2d$ grid were also  investigated~\cite{Leonhardt:1995aa,Zak:2011aa}, 
and used to deal with different aspects of quantum
information~\cite{Miquel:2002aa,Paz:2002aa}}.

\sugg{Another important line of research has focused on a 
phase space is pictured as a $d \times d$
lattice. It was started by  Buot~\cite{Buot:1973aa},
who introduced a discrete Weyl transform that generates
a Wigner function on the toroidal lattice ${\mathbb{Z}}_d$. This is in
the same vein of the pioneer work of Schwinger~\cite{Schwinger:1960aa}, 
who clearly recognized that the expansion of arbitrary operators in terms of certain
operator basis was the crucial concept in setting a
proper phase-space description. Indeed, he identified
the finite counterpart of the Weyl-Heisenberg group,
which describes the canonical conjugacy position-momentum
and that can be used to define a $d \times d$ phase
space~\cite{Weil:1964aa}. More recently, these ideas have been rediscovered and
developed further by other
authors~\cite{Galetti:1988aa,Cohendet:1988aa,Galetti:1992aa,Kasperkovitz:1994aa,Rivas:1999aa,Vourdas:2004aa,Ligabo:2016aa}.} 

Actually, when the dimension $d$ is a power of a prime, points in the
$d \times d$ grid must be labeled with elements of the Galois field
$\Gal{d}$: only by doing this we can endow the phase space with
geometric properties \sugg{similar to those of} the ordinary plane. Note also
that though the restriction to powers of primes rules out many quantum
systems, this formulation is ideally suited for the \sugg{time-honored
example} of $N$ qubits we deal in this paper.

\sugg{These satisfactory geometrical properties are in the realm of
  the most popular approach to deal with the discrete Wigner function,
  which is due to Wootters~\cite{Wootters:1987aa,Gibbons:2004aa}.}
This leads to a non-unique procedure of \sugg{relating states} in the
Hilbert space with lines in the grid. Such a map exhibits an important
property inherited from the \sugg{continuous} case: the sum of the
Wigner function along the line associated with a given state gives the
probability distribution in this state. Furthermore, this construction
also satisfies all the \emph{bona fide} requirements: invertibility,
Hermiticity, normalization and covariance under discrete displacements
generated by the Pauli group~\cite{Chuang:2000fk}.

On the other hand, \sugg{these} straight lines are
intimately related with the concept of mutually unbiased
bases~(MUBs)~\cite{Delsarte:1975aa,Wootters:1986aa}: eigenstates of
sets of $N$ commuting operators labelled with points of mutually
non-intersecting rays (lines passing through origin) determine MUBs.

\sugg{A complete set} of MUBs can be reduced to \sugg{an} arrangement
of $d^{2}-1$ disjoint operators into $d+1$ classes each containing
$d-1$ commuting operators. Eigenstates of lines in such a table with
$(d-1) \times(d+1)$ entries form
MUBs~\cite{Bandyopadhyay:2002aa}. Interestingly, these operators can
be organized in several nontrivial tables, leading to different
factorization properties~\cite{Romero:2005aa}. Here, we are interested
only in unitary equivalent sets of MUBs. It has been
noticed~\cite{Bjork:2007aa} that such arrangements are related with
special types of geometric structures in the discrete phase space, the
so-called commutative curves. A bundle of $d+1$ non-intersecting
curves determines the set. In principle, to each of
these bundles one can link a Wigner function with all the
required properties, in such a way that the traditional Wootters
approach is recovered for the special case of rays.  Obviously, to a
given state correspond different Wigner functions based on different
MUBs and the suitable choice of these MUBs depends on the entanglement
structure of the state.

In this paper, we go one step further and provide an explicit form of
phase-point operators \sugg{for $N$ qubits} corresponding to MUBs with
different factorization structures. \sugg{It results that these
  kernels are not equivalent under transformations connecting
  different sets of MUBs, but preserve the basic tomographic property,
  allowing to express the Wigner function of any state as a linear
  combination of measured probabilities. In addition, the Clifford
inequivalence leads to the possibility of finding non-stabilizer
states with non-negative Wigner functions, which clashes with
  previous results for the discrete
  case}~\cite{Galvao:2005aa,Cormick:2006aa,Gross:2006aa}. 

\section{Curves in phase space}

\sugg{For a system of $N$ qubits, the} Hilbert space is
the tensor product $\mathbb{C}^{2}\otimes \cdots \otimes
\mathbb{C}^{2}=\mathbb{C}^{2^{n}}$. Let $|k_{1},\ldots ,k_{N}\rangle $
($k_{i}\in \mathbb{Z}_{2}$) an orthonormal basis in
$\mathbb{C}^{2^{N}}$. We can label this basis by
$\kappa \in \mathbb{F}_{2^{N}}$, so that
\begin{equation}
  \kappa =\sum_{i=1}^{N}k_{i}\,\theta_{i}\,,  \label{alpha}
\end{equation}
where $\{\theta_{1},\ldots ,\theta_{N}\}$ is a self-dual basis [i.e., $
\tr (\theta_{i}\,\theta_{j})=\delta_{ij}$, with $\tr (\alpha
)=\alpha +\alpha ^{2}+\ldots +\alpha^{2^{N-1}}$, and  
$\alpha \in \mathbb{F}_{2^{N}}$].

The generators of the Pauli group $\mathcal{P}_{N}$ \sugg{are}
\begin{equation}
Z_{\alpha}=\sum_{\kappa}\chi (\kappa \alpha )\,|\kappa \rangle \langle
\kappa |\,,\qquad X_{\beta}=\sum_{\kappa}|\kappa +\beta \rangle \langle
\kappa |\,  \label{XZgf}
\end{equation}
and satisfy the commutation relations $Z_{\alpha}X_{\beta}= 
\chi (\alpha \beta )\,X_{\beta}Z_{\alpha}$, with $\chi (\alpha )=\exp [i\pi 
\tr (\alpha )]$ \sugg{being an} additive character. In
addition, $Z_{\alpha}$ and $X_{\beta}$ are related through the finite
Fourier transform~\cite{Klimov:2005aa}. 
 
\sugg{The operators (\ref{XZgf}) can be factorized in the form}
\begin{equation}
  Z_{\alpha}=
  \sigma_{z}^{a_{1}}\otimes \cdots \otimes \sigma_{z}^{a_{N}} ,
  \qquad 
  X_{\beta}=\sigma_{x}^{b_{1}}\otimes \cdots \otimes \sigma_{x}^{b_{N}}\,,
\end{equation}
where $a_{i}=\tr(\alpha \theta_{i})$ and
$b_{i}= \tr(\beta \theta_{i})$ correspond to the expansion
coefficients for $\alpha $ and $\beta $ in the self-dual basis.

The phase space can be appropriately labeled by the discrete points
$ (\alpha, \beta)$~\cite{Wootters:2004aa}, which are precisely the
indices of the operators $Z_{\alpha}$ and $X_{\beta}$: $\alpha$ is the
``horizontal'' axis and $\beta$ the ``vertical'' one.

\sugg{A stabilizer state is a simultaneous eigenvector of a maximal
  set of commuting observables in the Pauli group. A complete set of 
 stabilizers is given by} a set of $2^{N}$ disjoint
commuting monomials $\{Z_{\alpha (\tau )}X_{\beta (\tau )}\}$, expressed as 
\begin{equation}
\alpha (\tau )=\sum_{r=0}^{N-1}\alpha_{r}\,\tau ^{2^{r}} \,,
 \qquad 
\beta (\tau )=\sum_{r=0}^{N-1}\beta_{i}\,\tau ^{2^{r}}\,,   
\label{curve1}
\end{equation}
with $\alpha_{r},\beta_{r}\in \mathbb{F}_{2^{N}}$ and such that 
\begin{equation}
\sum_{r=0}^{N-1} \alpha_{p-r}^{2^{r}}  \beta_{q-r}^{2^{r}} = 
\sum_{r=0}^{N-1}\alpha_{q-r}^{2^{r}} \beta_{p-r}^{2^{r}} \, .
\end{equation}
We can look at these functions as curves $\Gamma =( \alpha (\tau), 
\beta (\tau )) $ in phase space. We impose that they pass through
the origin: $(\alpha (0),\beta (0))=(0,0)$; that is, $Z_{\alpha (0)}X_{\beta
(0)}=\openone$. We call them stabilizer curves. The disjointness
is in agreement with the fact that they have no self-intersections: all 
the $2^{N}$ pairs $(\alpha (\tau ),\beta (\tau ))$ are
different. \sugg{Consequently, to each stabilizer curve $\Gamma $ 
corresponds a basis $\{|\Psi_{\kappa}^{\Gamma} \rangle \}$, with $\kappa \in 
\mathbb{F}_{2^{N}}$.}

It follows from (\ref{curve1}) that summing the coordinates of any two
points of a stabilizer curve we obtain another point on the curve 
\begin{equation}
\alpha (\tau +\tau^{\prime}) = 
\alpha (\tau ) + \alpha (\tau ^{\prime}),
\qquad 
\beta (\tau + \tau ^{\prime}) = 
\beta (\tau )+\beta (\tau ^{\prime}) \, .  
\label{add}
\end{equation}
In other words, the stabilizers $\{Z_{\alpha (\tau )}X_{\beta (\tau )}\}$
form an Abelian group under multiplication, which is generated, e.g.,
by $\{Z_{\alpha (\theta_{i})}X_{\beta (\theta_{i})} \}$. 

A stabilizer curve is called regular when it can be represented in the
explicit form
\begin{equation}
  \beta =f(\alpha )\,,
  \quad \mathrm{or}\quad 
  \alpha =g(\beta )\,.  
  \label{RC}
\end{equation}
\sugg{Otherwise, the curve are called
  degenerate~\cite{Klimov:2009bk}. In that case, both $ \alpha $ and
  $\beta$ do not take some values in $\mathbb{F}_{2^{N}} $ and they
  are multivalued for some other values.}

The simplest form of stabilizer curves are the straight lines
\begin{equation}
  \alpha (\tau )=\mu \tau \,,\qquad \beta (\tau )=\nu \tau \,,
  \label{eq:sline}
\end{equation}
which can be represented in the regular form $\beta =\lambda \alpha $
(or $ \alpha =0$ for the vertical axis). It is a well established
result~\cite{Wootters:1989aa} that the operators
$\{Z_{\alpha}X_{\beta =\lambda \alpha}\}$ commute for any fixed
value of $\lambda \in \mathbb{F}_{2^{N}}$, while the eigenstates of
the set $\{Z_{\alpha}\}$ define the standard computational basis
$|\kappa \rangle $.

The regular curves can always be transformed into the horizontal [for curves 
$\beta =f(\alpha )$] or the vertical [for curves $\alpha =g(\beta )$] axes.
This can be accomplished by a pair of symplectic operations ($z$- and
$x$-rotations) such that  
\begin{equation}
P_{f}Z_{\alpha}P_{f}^{-1}\sim Z_{\alpha}X_{f(\alpha )}\,,
\qquad
Q_{g}X_{\beta}Q_{g}^{-1}\sim Z_{g(\beta )}X_{\beta}\,,  
\label{eq:PQ}
\end{equation}
the symbol $\sim $ indicating here equality except for a phase. Both $P_{f}$
and $Q_{g}$ are unitary operators, and can be written as 
\begin{equation}
P_{f}=\sum_{\kappa}c_{\kappa}^{(f)}\, |\widetilde{\kappa}\rangle 
\langle  \widetilde{\kappa}|\,,\qquad 
Q_{g}=\sum_{\kappa}c_{\kappa}^{(g)}\,|\kappa
\rangle \langle \kappa |\,,  
\label{V}
\end{equation}
where $|\widetilde{\kappa}\rangle $ are the eigenstates of
$X_{\beta}$.  The coefficients $c_{\lambda}^{(f)}$ satisfy the
recurrence relation
\begin{equation}
  c_{\kappa}^{(f)}\,c_{\kappa ^{\prime}}^{(f)}= 
  \chi  \left (  \kappa ^{\prime} f(\kappa ) \right ) 
  \,c_{\kappa +\kappa ^{\prime}}^{(f)}\,,
  \qquad 
  c_{0}^{(f)}=1\,,
  \label{eq:recrel}
\end{equation}
and analogously for $c^{(g)}$. In general, Eq. (\ref{eq:recrel})
admits multiple solutions, as discussed in detail in
Ref.~\cite{Klimov:2006aa}.  Thus, given a curve we can immediately
obtain the eigenstates of the set of commuting monomials attached
to this curve: for instance,
$|\Psi_{\kappa}^{\Gamma =f(\alpha )}\rangle =P_{f}|\kappa \rangle $.

\sugg{One of the most fundamental characteristics  of a stabilizer
  curve $\Gamma =\{Z_{\alpha (\tau )}X_{\beta (\tau )}\}$  is its factorization
structure; that is, the possibility of  parsing each monomial $Z_{\alpha
}X_{\beta}$  into smaller mutually commuting subsets
containing $1\leq k\leq N$ single-qubit operators:
\begin{equation}
\fact ( \Gamma )= 
\{m_{1},m_{2},\ldots ,m_{N}\} \,,  
\label{1_curve_part}
\end{equation}
where $\quad 0<m_{1}\leq m_{2}\leq \ldots \leq m_{N}$ ($m_{j}\in N$)
is the number of particles in the $j$-th block that cannot be
factorized into commuting sub-blocks anymore. It is 
clear that $\{m_{1},m_{2},\ldots ,m_{N}\}$  is just a partition of
the integer $N$, so the maximum number of terms is $N$, which
corresponds to a completely factorized curve, $\fact ( \Gamma  )= 
\underbrace{\{1,1,\ldots ,1\}}_{N}$, and the minimum number of terms
is one, for a completely non-factorized curve $\fact(
\Gamma )= \{N\}$.}

\section{Mutually unbiased bases from curves}

\sugg{The bases related} to nonintersecting curves $\Gamma $
and $\Gamma ^{\prime}$ are unbiased~\cite{Klimov:2009bk}; that is, 
\begin{equation}
|\langle \Psi_{\kappa}^{\Gamma}|
\Psi_{\kappa ^{\prime}}^{\Gamma^{\prime}}\rangle |^{2} = 
\frac{1}{2^{N}} \,, 
 \label{eq:unbias}
\end{equation}
so that a bundle of $2^{N}+1$ mutually nonintersecting curves define a
complete set of MUBs.

The simplest bundle is formed by the rays $\{\beta =\lambda \alpha , \alpha
=0\}$. The corresponding (standard) set of MUBs will be denoted as
$\{|\Psi_{\lambda ;\kappa}\rangle ,|\tilde{\Psi}_{\kappa}\rangle
\}$, where $|\tilde{\Psi}_{\lambda}\rangle $ are the eigenstates of $X_{\beta}$. The
set $\{|\Psi_{\lambda ;\kappa}\rangle \}$ is constructed as
$|\Psi_{\lambda ;\kappa}\rangle =P_{f=\lambda \alpha}|\kappa \rangle
$.  \sugg{This allows one to establish a canonical association
  between basis elements  and straight lines: 
\begin{equation}
|\Psi_{\lambda ;\kappa} \rangle \Longleftrightarrow 
\{\beta =\lambda \alpha +\kappa \} \, , 
\qquad
|\tilde{\Psi}_{\kappa}\rangle \Longleftrightarrow \{\alpha
=\kappa \},  \label{CA}
\end{equation}
where the ray $\beta =0$ is associated with the state $|\kappa =0\rangle $
(the only state with all positive eigenvalues), and the parallel lines $
\beta =\kappa $ correspond to the shifted states $|\kappa \rangle
=X_{\kappa}|0\rangle $.}  

\sugg{Our next  observation is that the ``rotated'' bases}
\begin{equation}
|\Psi_{\lambda ;\kappa}^{(f,g,h)}\rangle =
P_{h}Q_{g}P_{f}|\Psi_{\lambda;\kappa}\rangle \,,
\qquad 
|\tilde{\Psi}_{\kappa}^{(f,g,h)}\rangle
=P_{h}Q_{g}P_{f}|\tilde{\Psi}_{\kappa}\rangle \,,  
\label{C_bases}
\end{equation}
preserve the mutually unbiasedness inherited from the standard set
$\{|\Psi_{\lambda ;\kappa}\rangle , |\tilde{\Psi}_{\kappa}\rangle
\}$, so that
\begin{equation}
|\langle \Psi_{\lambda ;\kappa}^{(f,g,h)}
|\Psi_{\lambda ^{\prime};\kappa^{\prime}}^{(f,g,h)}\rangle |^{2}=
\delta_{\lambda \lambda ^{\prime}}\delta_{\kappa \kappa ^{\prime}}+
\frac{1}{2^{N}}(1-\delta_{\lambda\lambda ^{\prime}}),
\qquad 
|\langle \tilde{\Psi}_{\kappa ^{\prime}}^{(f,g,h)}
|\Psi_{\lambda ;\kappa}^{(f,g,h)}\rangle |^{2}=\frac{1}{2^{N}} \, .
\end{equation}
\sugg{Accordingly, $|\Psi_{\lambda ;\kappa}^{(f,g,h)}\rangle $ are eigenstates of
commuting} sets $\{Z_{\alpha_{\lambda}(\tau )}X_{\beta_{\lambda}(\tau )}\}$ with  
\begin{eqnarray}
\alpha_{\lambda}(\tau ) & = & \tau +g(\lambda \tau )+ g(f(\tau )) \,,  
\nonumber \\
\beta_{\lambda}(\tau ) & = & \lambda \tau +f(\tau ) +
h(\tau )+h(g(\lambda \tau))+h(g(f(\tau ))) \,, 
 \label{mub curves2}
\end{eqnarray}
\sugg{and} $|\tilde{\Psi}_{\kappa}^{(f,g,h)}\rangle $ are eigenstates of
$Z_{g(\tau )}X_{\kappa +h(g(\tau ))}$. \sugg{Therefore, for any fixed
  $\lambda$, the eigenstates of  the set 
$\{Z_{\alpha_{\lambda}(\tau )}X_{\beta_{\lambda}(\tau )}\}$ can be
associated with $2^{N}$  
mutually nonintersecting curves parallel to (\ref{mub curves2}),
whereas the eigenstates of $Z_{g(\tau )}X_{\tau+h(g(\tau ))}$ are
associated with curves parallel to  
\begin{equation}
\beta =\tau +h(g(\tau )) \, ,
\qquad 
\alpha =g(\tau ) \, .  
\label{mub curves0}
\end{equation}
Such sets of parallel curves, known as striations, have} the following
structure:

a.-- Curves parallel to $(\alpha_{\lambda}(\tau ),\beta_{\lambda}(\tau
)) $ are of the form 
\begin{eqnarray}
\alpha_{\lambda}(\tau ,\kappa ) &= &
\tau +g(\lambda \tau )+g(f(\tau))+g(\kappa )\,,  \nonumber \\
\beta_{\lambda}(\tau ,\kappa ) & = & 
\lambda \tau +f(\tau )+\kappa +h(\tau)+
h(g(\lambda \tau ))+h(g(f(\tau )))+h(g(\kappa ))\,.  
\label{c1}
\end{eqnarray}

b.-- Curves parallel to $\beta =\tau +h(g(\tau )),\alpha =g(\tau )$ are of
the form 
\begin{eqnarray}
\alpha (\tau ,\kappa ) &=&\kappa +g(\tau )+g(f(\kappa ))\,,  
\nonumber\\
\beta (\tau ,\kappa ) &=&\kappa +f(\kappa )+h(\kappa )+
h(g(\tau))+h(g(f(\kappa ))) \,.  
\label{c2}
\end{eqnarray}

\sugg{We conclude then that the bundle (\ref{mub curves2}) and
  (\ref{mub curves0}) is unitarily equivalent to the standard one
  formed by the rays $\{\beta =\lambda \alpha ,\alpha =0\}$.  
The advantage of the parametrization in
(\ref{c1}) and (\ref{c2})  is that it preserves  the same
association between states and curves as in (\ref{CA}); viz,}
\begin{eqnarray}
|\Psi_{\lambda ;\kappa}^{(f,g,h)}\rangle \Longleftrightarrow 
\Gamma_{\lambda ;\kappa}^{(f,g,h)}  & = &
\{\alpha_{\lambda}(\tau ,\kappa ),\beta_{\lambda}(\tau,\kappa )\}, 
\nonumber \\
|\tilde{\Psi}_{\kappa}^{(f,g,h)}\rangle \Longleftrightarrow
\Gamma_{\tilde{\kappa}}^{(f,g,h)} & = &
\{\alpha (\tau ,\kappa ),\beta (\tau,\kappa )\}.
\end{eqnarray}

\sugg{The resulting $\Gamma_{\lambda ;\kappa}^{(f,g,h)}$ and $\Gamma
_{\tilde{\kappa}}^{(f,g,h)}$ satisfy an important property: any pair of
curves crosses at a single point, much in the same way as straight lines.
This property is quite obvious for regular curves (\ref{RC}), but far
from trivial for degenerate curves.}

A bundle may contain curves with different factorizations
(\ref{1_curve_part} ). We characterize different bundles with a set of
numbers that indicate the number of completely factorized curves
($\underbrace{\{1,1,...,1\}}_{N}$ structure), completely factorized
except a single two-particle block (curves of the type
$\{\underbrace{1,1,...,1}_{N-2},2\}$), etc., until completely
nonfactorized curves~$\{N\}$. We thus assign to the bundle the set of
numbers
\begin{equation}
  (\ell_{1},\ell_{2},...,\ell_{p(N)}),\qquad \sum_{j}\ell_{j}=2^{N}+1,
\end{equation}
which indicate the number of curves factorized in $N$ one-dimensional
blocks, $\ell_{1}$; the number of curves factorized in $N-2$
one-dimensional blocks and one two-dimensional block, $\ell_{2}$;
etc, and $ p(N)$ is the number of partitions of an integer $n$. For
instance, the bundle of curves $\{\beta =\lambda \alpha ,\alpha =0\}$
has the structure (3,0,6) in the three-qubit case. Examples of another
bundles of curves corresponding to different type of factorizations of
complete set of MUBs can be found in Ref.~\cite{Klimov:2009bk}.

It is worth noting here that application of a set of three
transformations in (\ref{C_bases}), which is an analog to the Euler
decomposition in the discrete case~\cite{Duncan:2009aa}, is the most
general transformation that allows to obtain any curve bundle starting
from the rays $\{\beta =\lambda \alpha ,\alpha =0\}$. If two
transformations $Q_{g}P_{f}$ already produce an arbitrary curve from
any of the ray $\beta =\lambda \alpha $, the last
$ P_{h} $-transformation is required to obtain a generic (\sugg{in
  particular, a degenerate}) curve from the $x$-axis, $\alpha
=0$. However, not always all three transformations are needed to
generate a bundle with a given factorization structure as it will be
exemplified below.

\section{Wigner function on the curves}

According to Wootters original proposal~\cite{Wootters:1987aa} the
kernel of the discrete Wigner function (also called phase point
operator) is constructed as
\begin{equation}
  \hat{w} (\alpha ,\beta )=\sum_{\lambda}\hat{Q} (\lambda )-\openone \,,  
  \label{wW}
\end{equation}
where $Q(\lambda )$ is a projector linked with a line
$\beta =\lambda \alpha +\gamma $ passing through the point
$(\alpha ,\beta )$. The Wigner function for a state with density
operator $\varrho $ is then
\begin{equation}
  W_{\varrho} (\alpha ,\beta ) = \Tr [ \varrho 
 \hat{w}(\alpha ,\beta )] \,,  
\label{_J008}
\end{equation}
and it has the desired properties~\cite{Bjork:2008fk}.

\sugg{More explicitly, the kernel can be written down in terms of
  projectors on the standard MUBs, associated with rays, as follows}
\begin{equation}
\hat{w}(\alpha ,\beta )= |\tilde{\Psi}_{\alpha}\rangle 
\langle \tilde{\Psi}_{\alpha}| + 
\sum_{\lambda ,\gamma} 
\delta_{\beta ,\alpha \lambda +\gamma} \; 
|\Psi_{\lambda ;\gamma}\rangle \langle \Psi_{\lambda ;\gamma}| 
- \openone \,.  
\label{wl}
\end{equation}
In this way, the Wigner function of the state $|\Psi_{\lambda;\gamma} 
\rangle $ is just a straight line 
\begin{equation}
W_{|\Psi_{\lambda ;\gamma} \rangle} (\alpha ,\beta )=
\delta_{\beta,\lambda \alpha +\gamma}\,.  
\label{WT line}
\end{equation}

\sugg{The Wigner kernel for a complete bundle 
$\{\Gamma^{l}= (\alpha (\tau) = f^{l} ( \tau) , \beta ( \tau ) = 
g^{l} ( \tau) ) \}$ (with $l=1,...,2^{N}$) can be constructed in the 
same way as in (\ref{wW}). Indeed, let us denote by
\begin{equation}
\{ \Gamma_{\kappa}^{l}\}  = 
\{ \alpha_{\kappa} ( \tau) = f_{\kappa}^{l} ( \tau), 
\beta_{\kappa} ( \tau ) = g_{\kappa}^{l} ( \tau ) \} 
 \label{G}
\end{equation}
sets of parallel curves in the corresponding striations (i.e. the
curves $\Gamma_{\kappa}^{l}$, with $\kappa \in \Gal{2^{N}}$, do not
intersect for a fixed value of $l$) and
$\{|\Psi_{\kappa}^{\Gamma ^{l}}\rangle \equiv
|\Psi_{\kappa}^{l}\rangle$ are the associated states. Then, the Wigner
kernel $\hat{w}(\alpha ,\beta )$ can be jotted down exactly as in
(\ref{wW}):
\begin{equation}
  \hat{w} (\alpha ,\beta ) = \sum_{l=1}^{2^{N}+1} 
  \sum_{\kappa ,\tau \in \Gal{2^{N}}} 
  \delta_{\alpha ,f_{\kappa}^{l}( \tau)}
  \delta_{\beta,g_{\kappa}^{l} ( \tau)} 
  | \Psi_{\kappa}^{l} \rangle \langle \Psi_{\kappa}^{l}| -
  \openone \, .
  \label{wcc}
\end{equation}
This kernel satisfies the crucial tomographic property: summing the
Wigner function (\ref{_J008}) along any curve $\Gamma_{\kappa}^{l}$
from the set (\ref{G}) we obtain the probability of finding the system
in the state $|\Psi_{\kappa}^{l}\rangle $ associated with this curve:
\begin{equation}
\sum_{\alpha ,\beta}\sum_{\tau} 
\delta_{\alpha ,f_{\kappa}^{l} ( \tau )}
\delta_{\beta ,g_{\kappa}^{l} ( \tau)} 
\Tr [ \hat{\varrho} \hat{w}(\alpha ,\beta ) ] =
2^{N} \langle \Psi_{\kappa}^{l} | \hat{\varrho} \Psi_{\kappa}^{l} \rangle .
\end{equation}
As a direct consequence of (\ref{wcc}), we obtain that the
Wigner function of a state $| \Psi_{\kappa}^{l} \rangle $  has the form
of the corresponding curve in the discrete phase-space: 
\begin{equation}
W_{| \Psi_{\kappa}^{l} \rangle} =  \sum_{\tau}
\delta_{\alpha ,f_{\kappa}^{l} ( \tau )}
\delta_{\beta , g_{\kappa}^{l} ( \tau)} \, .
\label{Wfgh}
\end{equation}
As we have seen in previous Section, any extended bundle (that
includes all the striations) can be obtained form the standard
Wootters set of straight lines by unitary transformations. Then, for the
bases related with the curves (\ref{mub curves2}) one has
\begin{eqnarray}
\hat{w}(\alpha ,\beta ) & = &
|\tilde{\Psi}_{\alpha}^{(f,g,h)}\rangle 
\langle  \tilde{\Psi}_{\alpha}^{(f,g,h)}| +
\sum_{\lambda ,\tau ,\kappa}
\delta_{\alpha ,\tau +g(\lambda \tau )+g(f(\tau ))+g(\gamma )}
 \nonumber  \\
&\times &\ 
\delta_{\beta ,\lambda \tau +f(\tau )+\gamma +
h(\tau )+h(g(\lambda \tau ))+h(g(f(\tau )))+h(g(\gamma ))} 
|\Psi_{\lambda ;\kappa}^{(f,g,h)}\rangle 
\langle \Psi_{\lambda ;\kappa         }^{(f,g,h)}| -
\openone \,,  
\end{eqnarray}
which reduces to (\ref{wl}) when $f(x)=0$, $g(x)=0$, 
and  $h(x)=0$, and consequently, 
\begin{eqnarray}
\sum_{\tau}W_{\varrho}(\alpha_{\lambda}(\tau ,\kappa ),\beta_{\lambda
}(\tau ,\kappa )) &=&2^{N}\langle \Psi_{\lambda ;\kappa}^{(f,g,h)}|\varrho
|\Psi_{\lambda ;\kappa}^{(f,g,h)}\rangle \,,  \notag \\
&&  \label{tom cond} \\
\sum_{\kappa}W_{\varrho}(\alpha (\tau ,\kappa ),\beta (\tau ,\kappa ))
&=&2^{N}\langle \tilde{\Phi}_{\kappa}^{(f,g,h)}|\varrho |\tilde{\Psi}%
_{\kappa}^{(f,g,h)}\rangle \,,  \notag
\end{eqnarray}
where $\{ \alpha_{\lambda}(\tau ,\kappa ), 
\beta_{\lambda}(\tau,\kappa )\}$  and  $\{ \alpha (\tau ,\kappa ),
\beta (\tau ,\kappa )\}$  are defined in (\ref{c1}) and (\ref{c2}).}

\sugg{Observe, that by construction the kernel (\ref{wcc}) satisfies the
standard covariance (under discrete shifts) condition. Moreover, it  
cannot be obtained from (\ref{wl}) by a simple application of
unitaries $P$ and $Q$ that transform one curve bundle into another.  
In particular, a transformed kernel 
\begin{equation}
\hat{w}_{gfh} ( \alpha ,\beta ) = P_{h}Q_{g}P_{f} \, 
\hat{w} ( \alpha ,\beta ) \, 
P_{f}^{\dag}Q_{g}^{\dag}P_{h}^{\dag} \, ,  
\label{WU}
\end{equation}
does not satisfy the marginality (\ref{tom cond}). Instead, to get
the probabilities associated to $|\Psi_{\lambda ;\kappa}^{(f,g,h)}\rangle$, 
one should sum the Wigner function   (\ref{WU}) over the
straight lines $\beta = \lambda \alpha + \kappa $.}

\begin{figure}[tbp]
\begin{center}
\includegraphics[height=7cm]{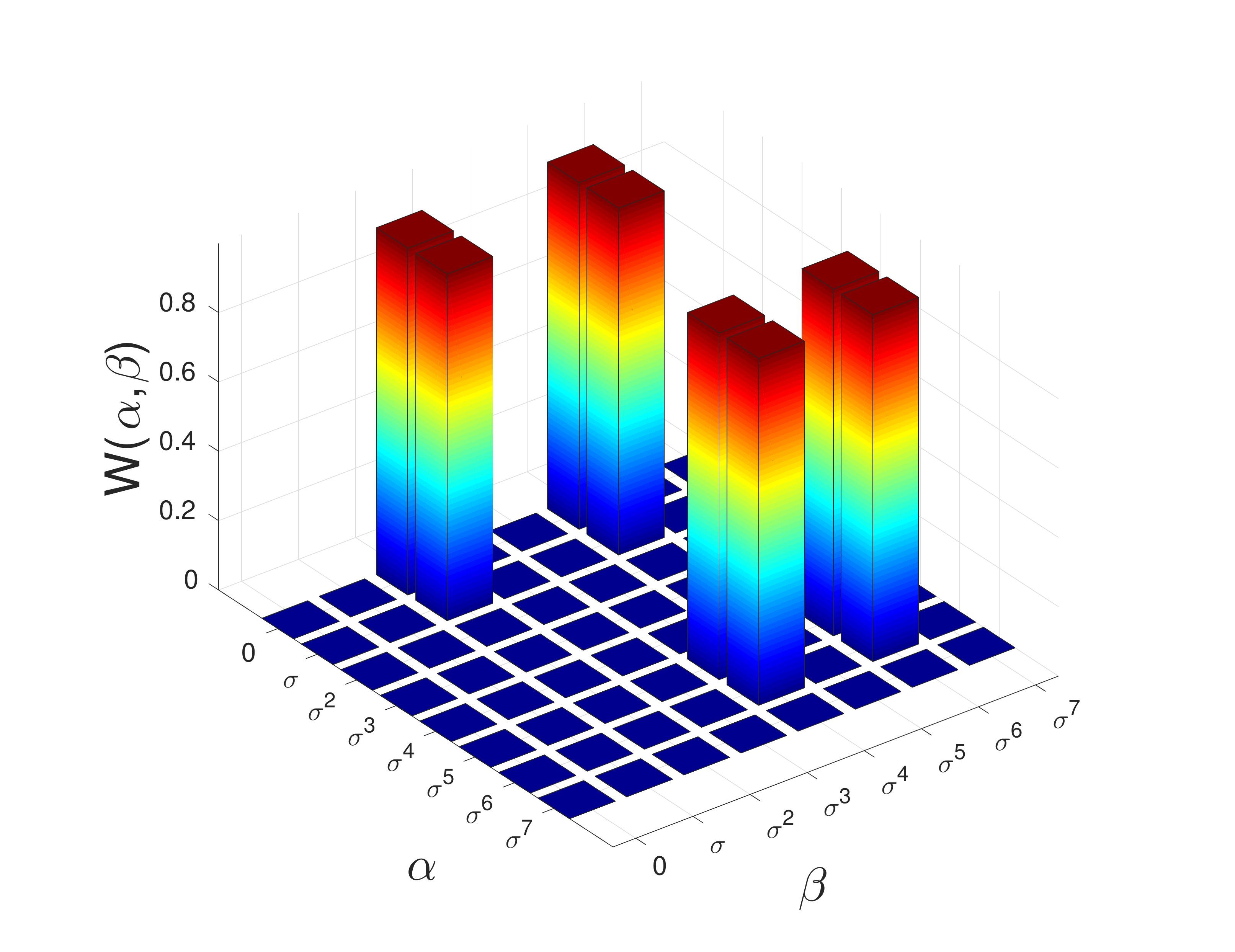}
\end{center}
\caption{Wigner function of an eigestate of a commuting set element of the
set (0,9,0) labeled by the curve $\alpha ={\sigma}^{5} +{\sigma}^{4}
\tau + {\tau}^{4}$, $\beta ={\sigma}^{4} {\tau}^{4}+ {\sigma}^{5}
{\tau}^{2} + {\sigma}^{4}$.}
\label{fig:stab}
\end{figure}

\begin{figure*}[tbp]
\centering
\subfigure{\includegraphics[width=0.49\columnwidth]{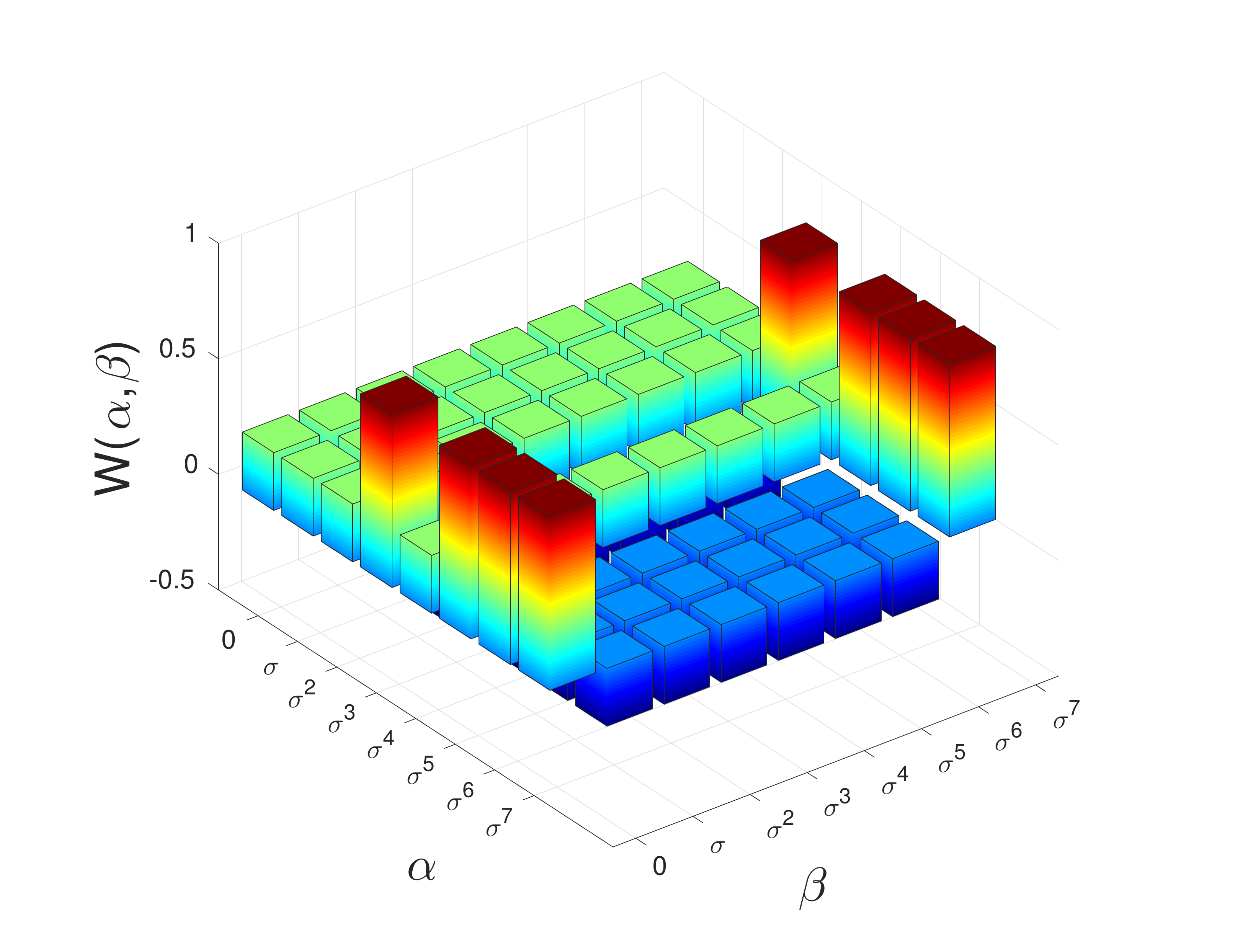}} %
\subfigure{\includegraphics[width=0.49\columnwidth]{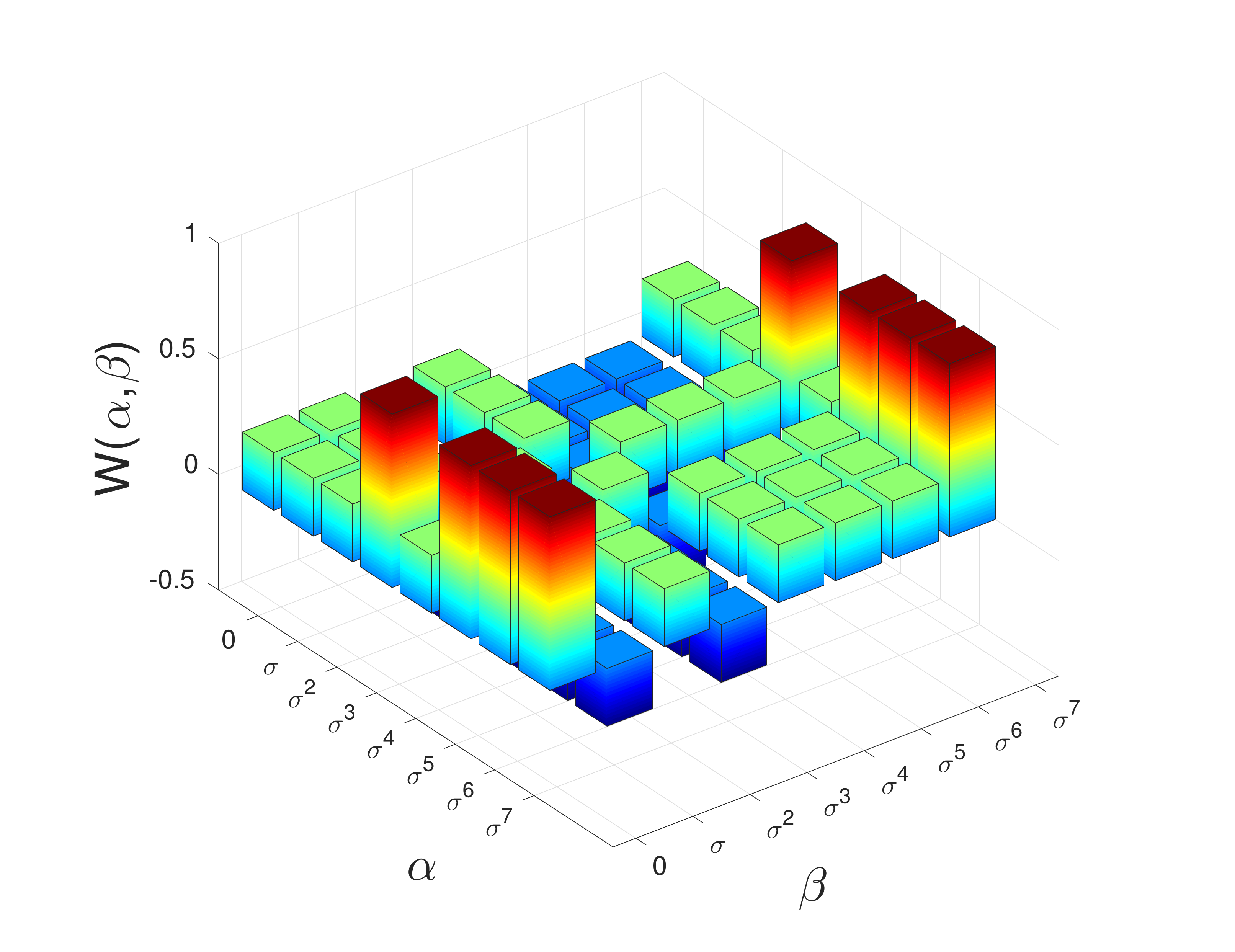}} %
\subfigure{\includegraphics[width=0.49\columnwidth]{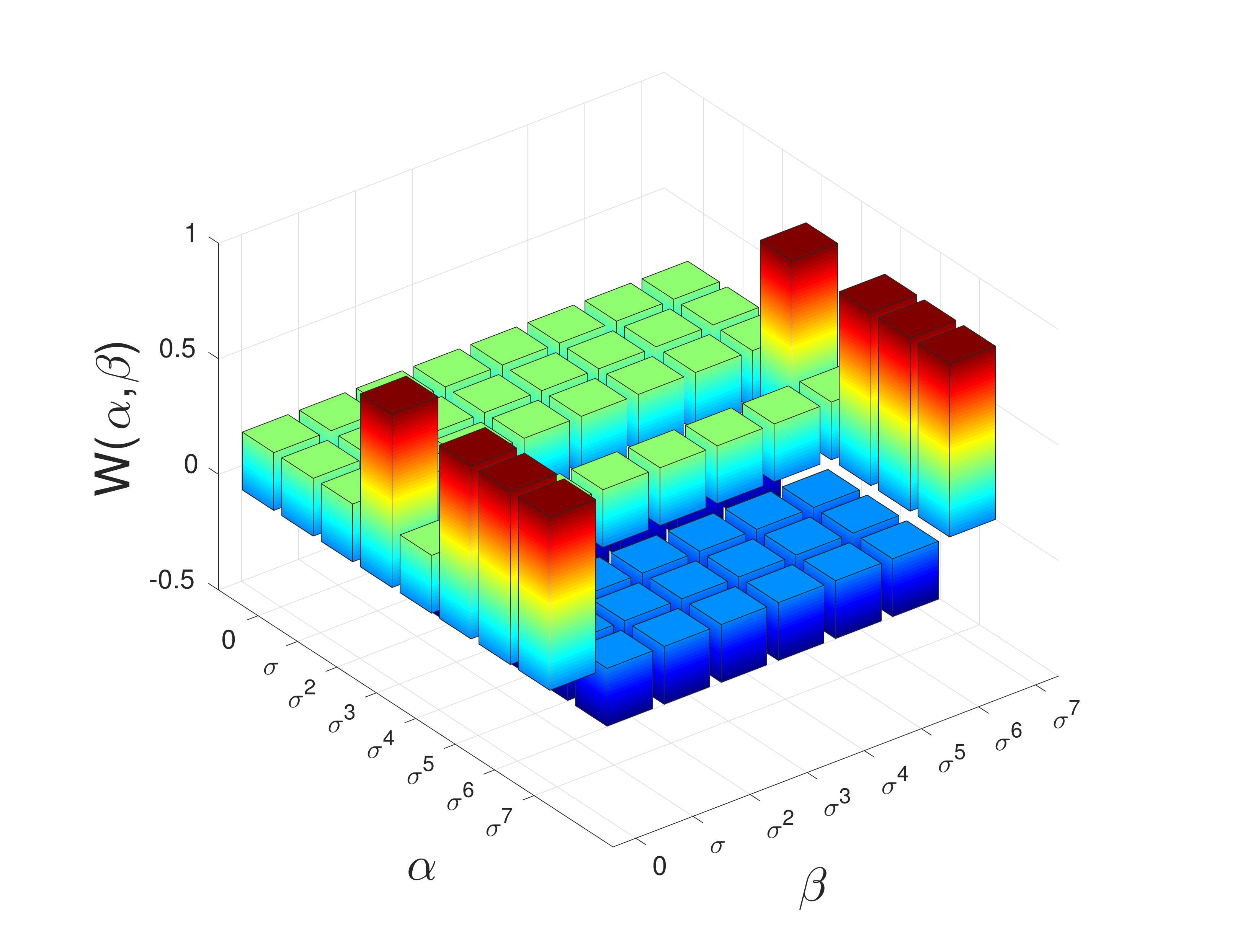}} %
\subfigure{\includegraphics[width=0.49\columnwidth]{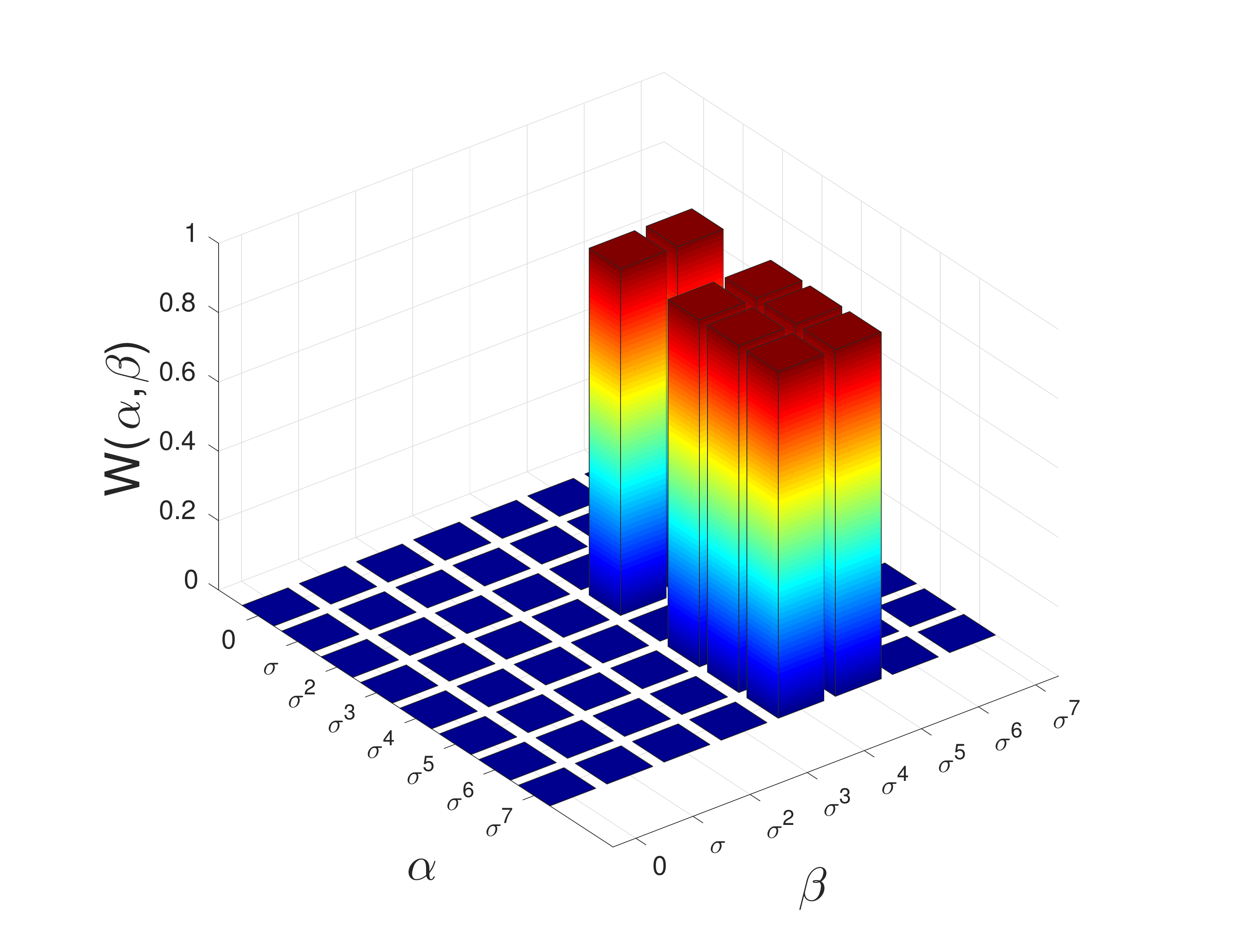}}
\caption{Wigner function for a GHZ state $(|000\rangle + |111\rangle)/%
\protect\sqrt{2}$ expressed in the four inequivalent sets of MUBs, with
factorization $(3,0,6)$ (top left), $(2,3,4)$ (top right), $(1,6,2)$ (bottom
left), and $(0,9,0)$ (bottom right).}
\label{fig:GHZ}
\end{figure*}

To exemplify this approach we consider the case of three qubits, for
which we know that the are four different sets of MUBs, with
factorizations $ (3,0,6)$, $(2,3,4)$, $(1,6,2)$, and $(0,9,0)$. The
standard set, as discussed before in relation with rays, is the
$(3,0,6)$. The MUBs with factorization $(1,6,2)$ can be obtained from
the standard one with the transformation $P_{f}$, with the curve
$f(\alpha )=\alpha +\alpha ^{2}+\alpha ^{4}$. The set with the
factorization $(2,3,4)$ requires application of two transformations
$P_{f}$ and $Q_{f}$, generated by the single curve
$f(\alpha )=\mu \alpha +\alpha ^{2}+\alpha ^{4}$ with any
$\mu \neq 0$. Finally, to generate the set with the factorization
$(0,9,0)$ a set of three transformations $P_{f}Q_{f}P_{f}$ is
required, although still  one curve
$f(\alpha )=\alpha +\sigma ^{2}\alpha ^{2}+\sigma \alpha ^{4}$ is
sufficient. Here, $\sigma $ is a primitive element of
$\mathbb{F}_{2^{3}}$, a root of $\sigma ^{3}+\sigma +1=0$.

In Fig.~\ref{fig:stab} we plot the Wigner function, for 
the MUB with factorization (0,9,0), of the eigenstate with all positive
eigenvalues of the stabilizer (degenerate) curve
$\alpha ={\sigma}^{5}+{\sigma}^{4}\tau +{\tau}^{4}$,
$\beta ={\sigma}^{4}{\tau}^{4}+{\sigma}^{5}{\tau}^{2}+{\sigma}^{4}$
(actually this curve is obtained by the transformation $%
P_{f}Q_{f}P_{f}$, $f(\alpha )=\alpha +\sigma ^{2}\alpha ^{2}+
 \sigma \alpha ^{4}$ from
the straight line $\beta ={\sigma}^{2}{\alpha}+{\sigma}^{2}$).
\sugg{For an explicit construction of the operators (\ref{eq:PQ}) we
  take here the particular solution of the recurrence
  (\ref{eq:recrel}) where the first $N$ coefficients are chosen
  positive $c_{\theta_{i}}^{(f)}=+\sqrt{\chi \left (\theta_{i}f(\theta_{i})
    \right ) }$. One can check the degeneracy of the corresponding
  curve.}

\sugg{The appearance of a quantum state may be very different under
  Wigner maps linked to MUBs with different factorizations.} In
Fig.~\ref{fig:GHZ} we plot the Wigner functions of a three-qubit
Greenberger-Horne-Zeilinger (GHZ) state
$(|000\rangle +|111\rangle )/\sqrt{2}$ for all possible
factorizations.  \sugg{For the factorization (0,9,0), the Wigner
  function contains only 8 points [although they do not form a curve,
  since (\ref{add}) is violated], while for the other factorizations
  it has a form of a ``real'' distribution, spread over all the phase
  space. This suggests that some factorizations could be more
  appropriate for representation of states, with particular
  correlation properties, than others.}

\begin{figure}[tbp]
\begin{center}
\includegraphics[height=7cm]{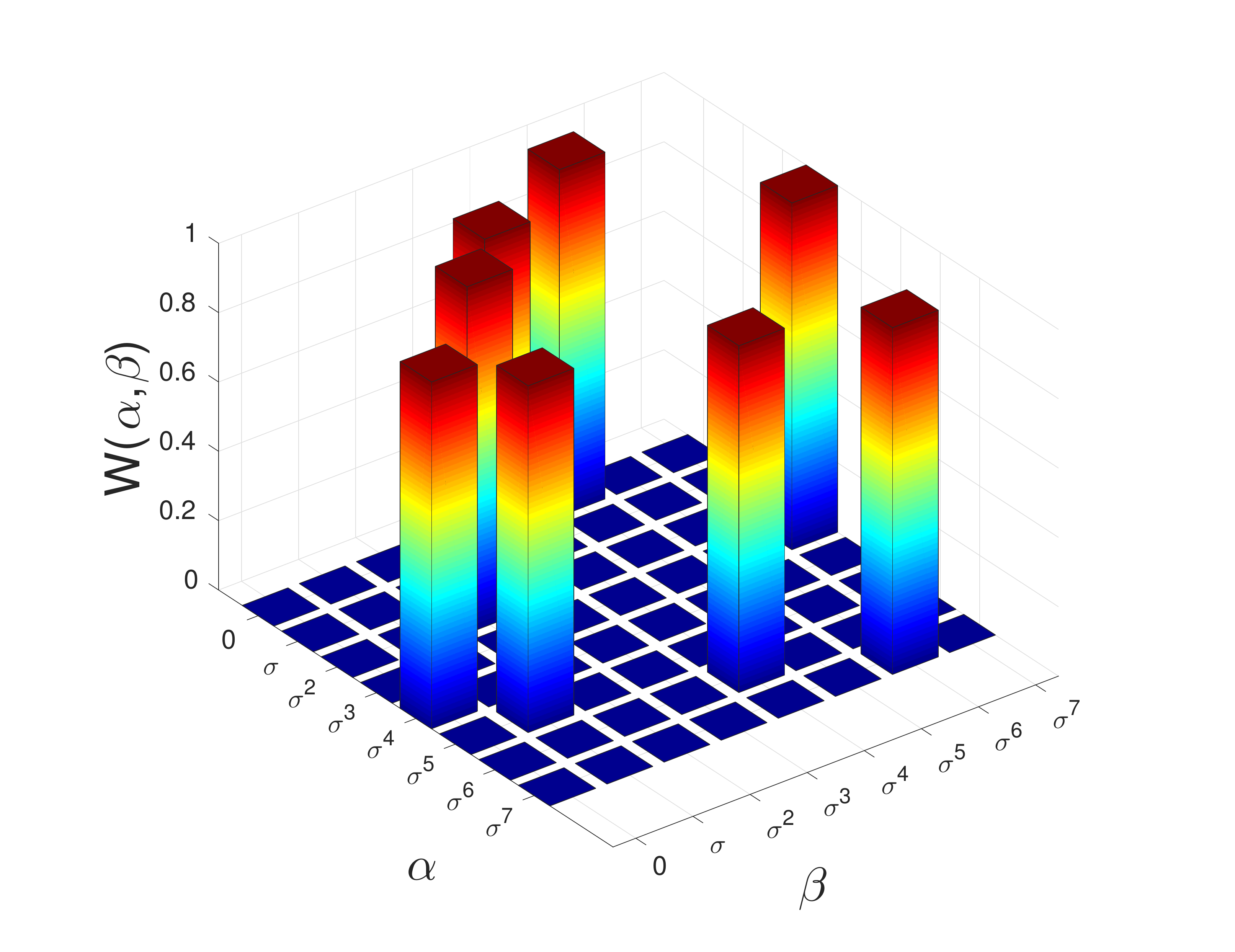}
\end{center}
\caption{Wigner function of the state $|\Psi_{\protect\lambda;0}\rangle =
P_{f= \lambda \alpha} | 0 \rangle$ in the set (0,9,0).}
\label{fig:rayo}
\end{figure}

\sugg{The Clifford inequivalence of the kernels (\ref{wl}) and
  (\ref{wcc}) brings about an unforeseen consequence: the 
possibility of finding non stabilizer states with positive Wigner
functions. In this respect, we recall that, in the standard Wootters
construction, corresponding to the set of MUBs $(3,0,6)$, the only states
with positive Wigner functions are stabilizer states
\cite{Galvao:2005aa,Cormick:2006aa,Gross:2006aa}. Indeed,
these states can be seen as the discrete counterparts of Gaussian for
continuous variable systems~\cite{Hudson:1974aa,Soto:1983aa} and the
negativity of the Wigner function as a measure of quantum 
correlations~\cite{Kenfack:2004aa,Siyouri:2016aa}.} 

As an example, one can show that the Wigner function of eigentstate $|\Psi
_{\lambda ;0}\rangle $ (with all positive eigenvalues) of the commuting set
labelled with points of the ray $\beta =\lambda \alpha $ in the set $(0,9,0)$
has a form of a line 
\begin{equation}
 W_{|\Psi_{\sigma ; 0}\rangle}^{(f,f,f)}(\alpha ,\beta ) =
\delta_{\beta,\lambda \alpha +\lambda ^{5}},
\qquad 
f(\alpha )=\alpha +\sigma ^{2}\alpha^{2}+\sigma \alpha ^{4}.
\end{equation}
This can be clearly observed in Fig.~\ref{fig:rayo}. Actually, eigenstates
of all the non-factorized rays $\beta =\lambda \alpha $ ($\lambda \neq 0,1$)
are represented by positive Wigner distributions in this set. Observe, that
such states, being completely non-factorized, are not eigenstates of any
stabilizer set in the set (0,9,0), which contains only bi-factorized bases.

This property strongly depends on the set of MUBs. For instance the
state $|\Psi_{\lambda ;0}\rangle $ is represented as a positive distribution
\sugg{(actually as a non-degenerate curve) in the set (1,6,2), whereas in the set
(2,3,4)  the same state has a complicated distribution.}

\section{Concluding remarks}

In summary, what we have shown is that for each complete set of MUBs, one
can construct a discrete Wigner map following the original Wootters idea:
the transformation kernel at a given point is obtained as a sum of
projectors on the basis states corresponding to the curves (associated with
such states) passing through this point. This construction generalizes the
standard one based on rays and cannot be obtained by a unitary
transformation of the former map. As an immediate consequence, we obtain
that the images of the basis states are not straight lines anymore, but some
specific curves in the phase space.

In addition, it appears that Wigner functions based on certain set of MUBs
may possess properties drastically different to the standard Wootters
construction. In particular positive distributions not necessarily
correspond to the stabilizer states.

\sugg{In principle, it would be interesting to extend these notions to
the continuous case.  However, this would require know the limit of $d
\rightarrow \infty$ of the MUBs. Although, this limit passing through
prime dimensions suggests the existence of an unlimited number of
MUBs, the question involve some subtle open questions~\cite{Weigert:2008aa}.}

\begin{acknowledgements}
  This work is partially supported by the Grant 254127  of CONACyT
(Mexico). L.~L.~S.~S. acknowledges the support of the Spanish MINECO
(Grant FIS2015-67963-P).
\end{acknowledgements}


\end{document}